# Exceptional electronic transport and quantum oscillations in thin bismuth crystals grown inside van der Waals materials

*Laisi Chen\*[1], Amy X. Wu\*[1], Naol Tulu[1], Joshua Wang[1], Adrian Juanson[2], Kenji Watanabe[3], Takashi Taniguchi[4], Michael T. Pettes[5], Marshall Campbell[1,5], Chaitanya A. Gadre[1], Yinong Zhou[1], Hangman Chen[6], Penghui Cao[6], Luis A. Jauregui[1], Ruqian Wu[1], Xiaoqing Pan[1,7,8], Javier D. Sanchez-Yamagishi[1]*

*1 - Department of Physics and Astronomy, University of California, Irvine, Irvine, CA, USA*
*2 - Department of Physics and Astronomy, California State University Long Beach, Long Beach, CA, USA*
*3 - Research Center for Functional Materials, National Institute for Materials Science, 1-1 Namiki, Tsukuba, Japan*
*4 - International Center for Materials Nanoarchitectonics, National Institute for Materials Science, 1-1 Namiki, Tsukuba, Japan*
*5 - Center for Integrated Nanotechnologies (CINT), Materials Physics and Applications Division, Los Alamos National Laboratory, Los Alamos, NM, USA*
*6 - Department of Mechanical and Aerospace Engineering, University of California, Irvine, Irvine, CA, USA*
*7 - Department of Materials Science and Engineering, University of California, Irvine, Irvine, CA, USA*
*8 - Irvine Materials Research Institute, University of California, Irvine, Irvine, CA, USA*

*\*these authors contributed equally to this work*

## Abstract

Confining materials to two-dimensional forms changes the behavior of electrons and enables new devices. However, most materials are challenging to produce as uniform thin crystals. Here, we present a new synthesis approach where crystals are grown in a nanoscale mold defined by atomically-flat van der Waals (vdW) materials. By heating and compressing bismuth in a vdW mold made of hexagonal boron nitride (hBN), we grow ultraflat bismuth crystals less than 10 nanometers thick. Due to quantum confinement, the bismuth bulk states are gapped, isolating intrinsic Rashba surface states for transport studies. The vdW-molded bismuth shows exceptional electronic transport, enabling the observation of Shubnikov–de Haas quantum oscillations originating from the (111) surface state Landau levels, which have eluded previous studies. By measuring the gate-dependent magnetoresistance, we observe multi-carrier quantum oscillations and Landau level splitting, with features originating from both the top and bottom surfaces. Our vdW-mold growth technique establishes a platform for electronic studies and control of bismuth's Rashba surface states and topological boundary modes. Beyond bismuth, the vdW-molding approach provides a low-cost way to synthesize ultrathin crystals and directly integrate them into a vdW heterostructure.

## Main Text

Our understanding of two-dimensional (2D) electronic physics has been significantly advanced by studying ultrathin van der Waals materials isolated by mechanical exfoliation[1]. However, this approach is generally inapplicable to other materials. In some cases, 2D growth of non-vdW materials is possible by molecular beam epitaxy (MBE) and other deposition techniques but often results in irregular surfaces or undesirable substrate interactions[2–5]. We consider an alternative 2D synthesis approach, where crystals are grown in the confined space between the layers of a vdW material. In confined growth, a mold defines the crystal geometry[6–8], but any surface roughness will be imprinted on the crystal and adversely affect electronic properties. Recently, it has been demonstrated that vdW materials can define atomically-smooth channels of nanoscale thickness[9]. Such channels are difficult to achieve via any other technique and are attractive molds for confined crystal growth.

Here, we demonstrate confined growth of ultraflat bismuth between layers of hexagonal boron nitride (hBN), a vdW material. Bismuth has played a key role in the development of quantum electronic physics due to its low carrier density, small effective mass, and the ability to grow pristine bulk crystals[10,11]. Recently, the boundaries of bismuth have received increasing attention due to the spin-momentum locking in 2D Rashba surface states and 1D helical edge modes[10,12,13]. These boundary modes have been probed primarily by STM and ARPES, which have revealed diverse phenomena[2,12,14]. However, transport studies have been limited by disorder in thin crystals where bulk conduction is gapped due to confinement. Via our vdW-mold technique, we produce crystals that enable intrinsic transport studies of thin bismuth and its surface states.

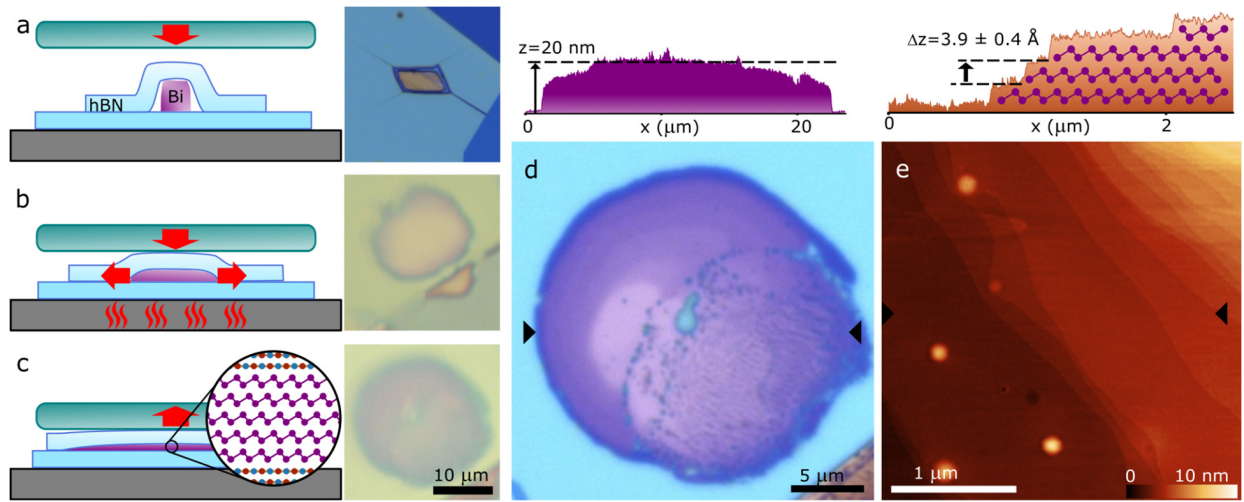

**Figure 1. Growth of ultrathin bismuth crystals inside a vdW-mold.** a-c, Cross-sectional schematics of the vdW-mold process with corresponding optical images of the bismuth. a, Bismuth flake encapsulated in hBN on a bottom substrate Si/SiO$_2$ before squeezing. b, Uniaxial compression (red arrow) is applied to the stack by a rigid top substrate (glass or sapphire) while the stage is heated. When the bismuth reaches its melting point, it rapidly compresses and expands laterally. c, Bismuth is cooled below its melting point and then pressure is removed, resulting in an ultrathin bismuth crystal. d, Optical image of the encapsulated vdW-molded bismuth (sample M30), black triangles indicate location of the AFM line trace (top) of the bismuth taken after removing the top hBN flake. This bismuth varies from 10 to 20 nm thick. e, AFM topography of the vdW-molded bismuth after removing the top hBN showing wide flat terraces. Black triangles show location of the line trace (top). The average step height is 3.9 ± 0.4 Å.

Our process for growing crystals within a vdW-mold is described in Figure 1. First, a micron-sized flake of bismuth is encapsulated in thin hBN layers using standard vdW transfer techniques[15] (SI Section 1). Next, the bismuth-hBN stack is compressed between two substrates, and then sequentially heated and cooled to melt and re-solidify the bismuth (Fig 1b). The process is performed inside an inert-gas glove bag to prevent oxidation. On melting, the liquid bismuth rapidly spreads out between the hBN layers and reduces in thickness due to the applied pressure. The squeezed form is retained when cooling into the solid phase and releasing the pressure, resulting in a thin bismuth crystal encapsulated in hBN (Fig 1c).

By squeezing the bismuth during the melt-growth, we reduced 250-500 nm thick flakes to ultrathin crystals ranging from 5 to 30 nm thick. Optically, the crystals exhibit large smooth areas with step-like contrast changes (Fig 1d and Fig S2). These smooth areas emerge from a rougher material centered at the location of the initial bismuth flakes, with the latter arising from a combination of small voids in the crystal, due to trapped gas within the vdW mold, as well as from the bismuth oxide that surrounded the original seed flake. After removing the top hBN layer, we characterize the bismuth surface by atomic force microscopy (AFM) (Fig 1e). The optically smooth regions correspond to flat terraces 0.5 to 5 μm wide separated by uniform steps. The average step height in Figure 1e is 3.9 ± 0.4 Å, which matches the thickness of a buckled hexagonal layer of bismuth, also known as the (111) "bilayer"[10].

To characterize the flatness of the hBN-molded bismuth, we plot the distribution of height deviations across a wide terrace (Fig 2a&b). Within the terrace of the hBN-molded bismuth, the surface is ultraflat with an average root mean square deviation $\delta_{Bi-BN} = 0.11\ nm$ (Fig 2b). By comparison, bismuth that has been molded by thermally-grown SiO$_2$ exhibits a roughness of $\delta_{Bi-SiO_2} = 0.27\ nm$. We ascribe this difference to the intrinsically-flat structure of the vdW-layered hBN ($\delta_{BN} = 0.05\ nm$) as compared to the amorphous SiO2 ($\delta_{SiO_2} = 0.25\ nm$). We have tested other mold substrates, such as graphite, sapphire and mica, and observed that the mold surface roughness is imprinted on the squeezed bismuth (SI Section 4), with hBN producing the overall best results.

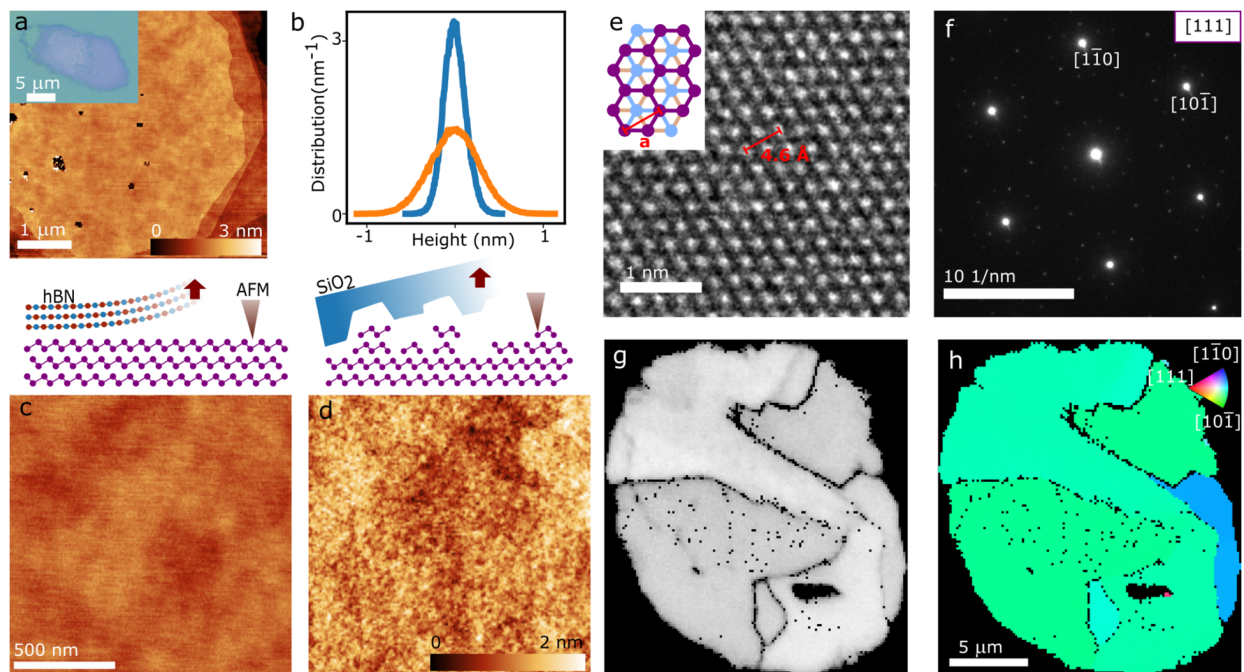

**Figure 2. Flatness and crystallinity of vdW-molded bismuth.** a, AFM topography of hBN-molded bismuth showing large flat terraces (sample M60). (Inset) Optical image of the sample. b, Height distribution of hBN-molded bismuth (blue) and SiO$_2$-molded bismuth (orange). c&d, AFM of the surfaces of hBN-molded bismuth (c) and SiO$_2$-molded bismuth (d). Top substrates are removed before AFM. Scale bar is the same for both images. The diagrams above depict the imprinting effect of the mold substrate on the bismuth. e, TEM real space image of hBN-molded bismuth. Inset, bismuth (111) lattice structure. f, TEM Selected area electron diffraction (SAED) of hBN-molded bismuth. g, Electron backscatter diffraction (EBSD) band contrast of the hBN-molded bismuth showing internal domain boundaries (dark lines). h, EBSD Y-axis inverse pole figure of the same sample showing overall (111) orientation with large domains differentiated by in-plane rotations.

To determine the structure and crystallinity of the vdW-molded bismuth, we performed transmission electron microscopy (TEM), electron backscatter diffraction (EBSD), and Raman spectroscopy (SI Section 5&7). All three measurements are consistent with the typical rhombohedral structure of bismuth. The TEM and EBSD diffraction measurements on 6 different samples reveal that the bismuth is highly crystalline and oriented along the (111) axis, corresponding to buckled honeycomb layers stacked parallel to the substrate (Fig 2e&f&h). The in-plane lattice constant measured from the real space image is 4.59 Å, which is within 1% of the bulk bismuth value 4.54 Å[10]. Diffraction measurements show single-crystal domains up to 10 microns in size, which are differentiated from other (111) domains by in-plane rotations (Fig 2h). The in-plane orientation of the bismuth shows no alignment with the encapsulating hBN layers. Additional structural characterization data can be found in the supplementary information.

By analyzing 20 different growths, we find that vdW-molded bismuth forms a dome shape with a typical thickness of 8 to 20 nm; terraces as thin as 5 nm can be observed near the crystal edge (Fig S4). The compression squeezes the bismuth until the top and bottom substrates elastically deform and contact around the bismuth-hBN stack, which we estimate to occur at 1 GPa pressure. Without compression, vdW-molding results in thicker bismuth crystals (t > 100 nm). This behavior is consistent with a simple continuum model, which predicts that thinner crystals are achievable by using more rigid substrates (Fig S12). Initial tests using sapphire as a top substrate instead of glass show 40% thinner bismuth on average (Fig S4). The continuum model is also qualitatively consistent with atomistic simulations, which show the synthesis of even sub-nanometer-thick 2D crystals at high compression (Fig S13).

To electrically characterize the vdW-molded bismuth, we fabricated Hall-bar shaped devices and measured them in a variable temperature cryostat (Fig 3). In all vdW-molded bismuth devices, we observe metallic behavior at low temperatures with a positive slope of resistance vs. temperature, dR/dT (Fig 3e and S15). The slope decreases with rising temperature, unlike the linear-T dependence observed in bulk bismuth, and is a signature of confinement effects in the vdW-molded bismuth[11]. In thin devices, with thicknesses t = 8 nm and 13 nm, the metallic dependence persists to room temperature, but in thicker flakes, t = 45 nm to 107 nm, dR/dT can become zero or negative above ~50 K, indicating an activated dependence. This behavior, measured in 6 devices, is well modeled by parallel conduction through a metallic channel and a semiconducting bulk that is gapped due to vertical confinement[16–18] (Fig S15). The metallic channel is identified with the (111) Rashba surface states, which are known from ARPES and STM studies to have electron and hole pockets[17,19,20]. This is consistent with our observation of a positive non-saturating magnetoresistance in all of our devices up to 12 T (Fig 3f & S16), indicating nearly compensated electron- and hole-like carriers.

The vdW-molded bismuth exhibits exceptional transport properties compared to MBE-grown thin films. Previously, thin bismuth studies have been characterized by weak temperature dependences with little metallic contribution and small residual resistance ratios (RRR) of 0.5 to 1.6[16,21–23]. By contrast, our vdW-molded devices exhibit strong metallic dependencies, showing that surface-derived states can dominate the conduction at room temperature. This results in substantially larger RRR values, such as 12 and 5.4 for the flat 8 nm and 13 nm devices, respectively (additional devices in Fig S15). The substantially larger RRR value indicates reduced scattering times, which is also consistent with our observation of 100x larger magnetoresistance signals than measured in MBE films (Fig 3f).

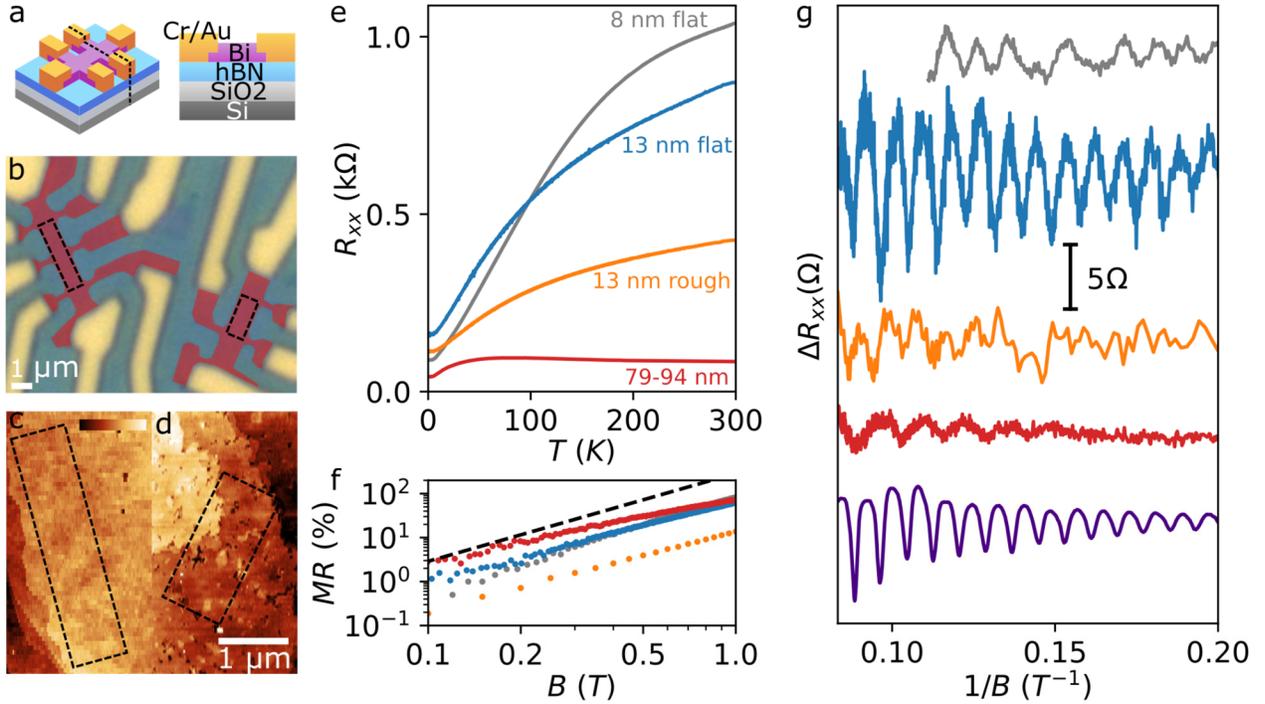

**Figure 3. Electronic transport and quantum oscillations in vdW-molded bismuth devices. a,** Isometric and cross-sectional schematics of the devices. **b,** False color optical image of the 13 nm thick vdW-molded bismuth devices, top hBN layer is removed. Bismuth is red, gold electrodes are yellow. **c&d,** Surface topography of the 13 nm thick bismuth devices, one that is ultraflat (c) (0.8 µm x 3.6 µm) and the other rough (d) (1.1 µm x 2.1 µm). Color scale range is 2.5 nm. **e,** Longitudinal resistance ($R_{xx}$) as a function of temperature for different thickness devices. **f,** Logarithmic plot of magnetoresistance as a function of out-of-plane magnetic field (T=1.6 K) for the same devices as in panel (e). Black dashed line is quadratic dependence. **g,** Quantum oscillations in $\Delta R(1/B)$, calculated by subtracting a smoothed background. Line colors correspond to the device labels in (e). Additional data from a 107 nm thick device is included (bottom-most purple line, scaled by 0.05x). Fermi surface area from top to bottom are 2.75e12, 2.82e12, 2.99e12, 1.99e12, and 2.60e12 cm$^{-2}$.

At high fields, quantum oscillations in the magnetoresistance emerge in the vdW-molded devices. The quantum oscillations are pervasive, occurring in 11 devices covering a wide range of thicknesses from 8 to 107 nm, with onset fields of 3 to 4 T (Fig 3g and SI Section 16). The dominant oscillations correspond to Fermi surface area of 2e12 to 3e12 cm$^{-2}$, which are 13-20x larger than that of bulk bismuth[24] and are consistent with STM and ARPES studies[17,19,20]. Such Shubnikov–de Haas oscillations in the magnetoresistance were famously first observed in bulk bismuth[25], but have not been previously observed in the bismuth surface states.

The presence of quantum oscillations in vdW-molded bismuth is striking, especially considering that the nanofabrication process exposes the bismuth to ambient air and polymers, which are avoided in MBE studies using *in-situ* measurements[16,22]. A key difference is that MBE-grown bismuth has irregular surfaces with dense steps, typically 20 to 100 nm apart[5,26], which can scatter electrons and provide parallel conduction[13,14]. By contrast, the vdW-molded bismuth typically has 10 to 100 times wider terraces (Fig S3). To study the effect of surface roughness on transport properties, we compare two devices from the same 13 nm thick bismuth flake: one fabricated on an ultraflat terrace and the other on a rougher region (Fig 3c&d). The rough device shows degraded transport relative to the flatter device, with reduced values for the RRR, magnetoresistance, and amplitude of quantum oscillation features (Fig 3e-g).

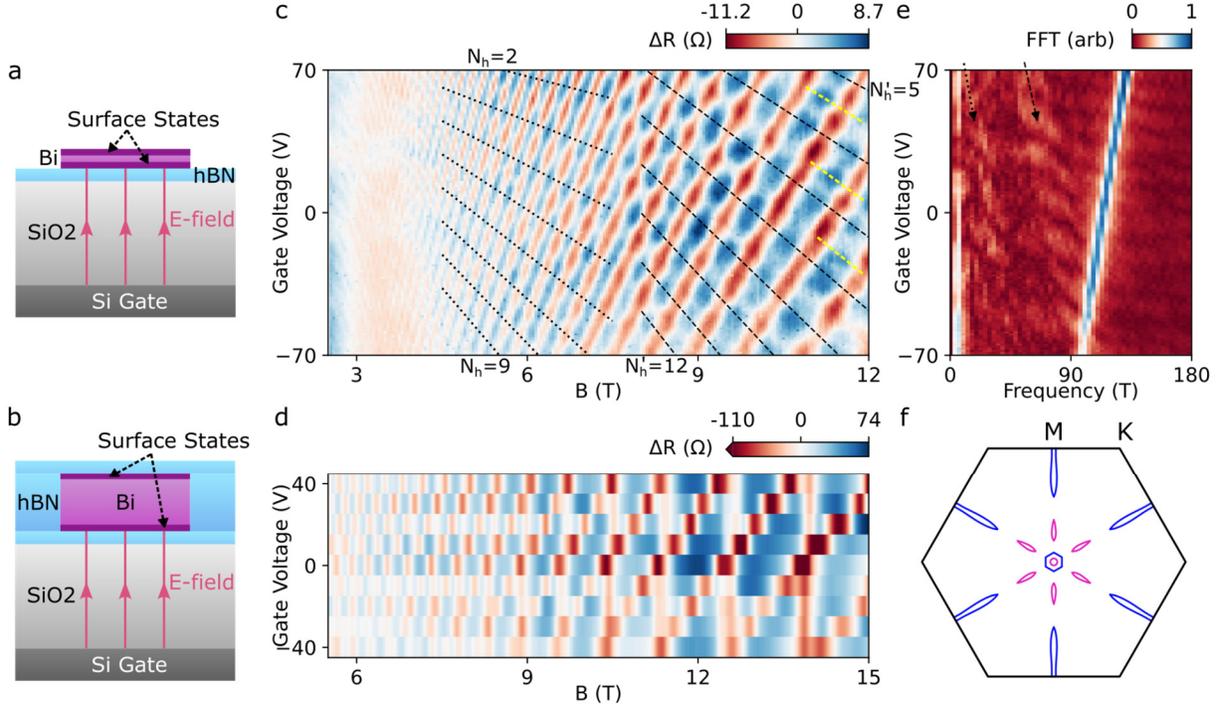

**Figure 4. Gate-dependence and surface state coupling in thin and thick vdW-molded bismuth. a**, Cross-sectional schematics of a 13 nm thick bismuth device showing similar gate capacitive coupling to the top and bottom surfaces. **b**, Cross-sectional schematics of a 107 nm thick bismuth device showing larger capacitive coupling of the gate to the bottom surface. **c&d**, Quantum oscillations measured as a function of gate voltage and magnetic field for the flat 13 nm device (c, T = 1.5 K) and the 107 nm device (d, T = 0.3 K). $\Delta R$ is calculated by subtracting a smooth background from magnetoresistance measurements. Landau fan fit is indicated by dashed lines based on (e). Devices are the same as in Figure 3. **e**, FFT of the quantum oscillations shown in c, showing one electron-like oscillation (blue streak feature) and two sets of hole-like oscillations indicated by the dashed arrow (same as in c). **f**, Fermi surface diagram of bismuth (111) surface states based on DFT calculations of 4.7 nm thick bismuth (Fig S10). Blue: electron pockets, purple: hole pockets.

The quantum oscillations in the vdW-molded bismuth disperse strongly with backgate voltage, resulting in clear Landau fan features (Fig 4). In the flat 13 nm device, two distinct fans are observed with opposite slopes in field vs. gate space, indicating electron-like and hole-like carriers (Fig 4c&e). At fields above 8 T, the hole-fan changes to a different slope. At zero gate voltage, the 1/B frequencies of the oscillations correspond to Fermi surface areas of 2.77e12 $cm^{-2}$, 0.73e12 $cm^{-2}$, and 1.94e12 $cm^{-2}$ for the electron and two hole pockets, respectively. Interestingly, each Fermi surface changes similarly over the 140V gate voltage range with $\Delta n$ = 0.78, 0.93, and 1.13e12 $cm^{-2}$. By fitting Landau fans to the datasets, we observe that the highest gate voltage depletes the low-density hole pocket to Landau index $N_h$ = 2. With a larger gate voltage, we would expect to reach the quantum limit and completely deplete the pocket. Such a large gate modulation is significant, as the different Fermi surfaces are expected to carry different spin textures due to Rashba spin-orbit coupling[10].

Previous theoretical and ARPES studies of the bismuth (111) surface states have identified three types of Fermi surfaces: a Gamma-point electron pocket, a 6-fold degenerate hole pocket, and a set of electron pockets near the M point (Fig 4f)[17,18]. Landau level features have only been observed by STM, and were ascribed to the Gamma electron pocket and the oblong hole valleys[19,20]. The pocket sizes we observe are comparable to these previous studies, but it is surprising that all the pockets exhibit similar gate couplings despite large expected differences in their degeneracies. The origin of the change in the hole pocket size at fields above 8 T is also unknown.

Transport is differentiated by measuring contributions from the top and bottom surfaces, while STM and ARPES probe only the top surface. Generically, asymmetries between the surfaces are expected both due to the device structure and the applied electric fields. In the 13 nm device, we do not observe any splitting induced by the backgate, suggesting that the surfaces are well coupled due to their proximity. Instead, we observe magnetic-field-induced splittings of the Landau levels at 11 T (Fig 4c, yellow lines), which can arise from exchange-induced nematicity as is observed in STM[20]. By contrast, in the 107 nm device (Fig 4d), we observe clear gate-induced splitting of the electron-pocket Landau fan, resulting in two sets of oscillations with applied gate voltage, one strongly gate-dependent and the other independent of gate. We ascribe these oscillations to the decoupled top and bottom surface states, where the bottom surface couples more strongly to the backgate and screens the electric field from reaching the top surface. This demonstrates how the surfaces of thin bismuth crystals intrinsically support independently-controllable 2D systems.

To summarize, we demonstrate a method to synthesize ultrathin and flat bismuth crystals within a nanoscale vdW-mold. The confined bismuth exhibits quantum oscillations of the magnetoresistance originating from the (111) surface states, which we can strongly modulate via the field effect. These bismuth surface bands are known to feature spin-momentum locking, and hence electrical currents can generate spin polarization as in conventional 3D topological insulators (TIs)[10,27–29]. Moreover, vdW-molded bismuth can shed light on the transport behavior of intrinsic helical edge modes, which have been observed in STM and ascribed to bismuth being a higher-order topological insulator[12].

Our mechanical modeling and atomistic simulations suggest that vdW-molding can offer a route to 2D bismuthene, which is predicted to be a large-gap 2D topological insulator[30,31]. Beyond bismuth, our initial tests show promise in applying the vdW-mold technique to other soft materials, such as gold, indium and tin. We anticipate these techniques will enable a new research approach combining ultrathin vdW and non-vdW materials.

## Acknowledgments

The fabrication and measurements of ultrathin bismuth devices was primarily supported by the Air Force Office of Scientific Research under award number FA9550-21-1-0165 and FA9550-23-1-0454. Materials characterization and technique development was supported by the National Science Foundation (NSF) Materials Research Science and Engineering Center (MRSEC) program through the UC Irvine Center for Complex and Active Materials (DMR-2011967) Seed Program. The authors acknowledge the use of facilities and instrumentation at the Integrated Nanosystems Research Facility (INRF) in the Samueli School of Engineering at the University of California Irvine and at the UC Irvine Materials Research Institute (IMRI), which is supported in part by the NSF MRSEC through the UC Irvine Center for Complex and Active Materials. Film deposition work was performed using instrumentation funded by DURIP award FA2386-14-1-3026. Raman spectroscopy was supported by the Laboratory Directed Research and Development program of Los Alamos National Laboratory under project number 20210782ER (M.T.P.,M.C.). A portion of this work was performed at the National High Magnetic Field Laboratory, which is supported by National Science Foundation Cooperative Agreement No. DMR-2128556 and the State of Florida. This work was performed, in part, at the Center for Integrated Nanotechnologies, an Office of Science User Facility operated for the U.S. Department of Energy (DOE) Office of Science. K.W. and T.T.

acknowledge support from the JSP KAKENHI  (Grant Numbers 19H05790, 20H00354 and 21H05233). The authors thank I. Krivorotov and A. Khan for the assistance and use of their sputtering machine. The authors thank V. Fatemi, A.F. Young, M.Q. Arguilla, and X. Yan for productive discussions, as well as the technical assistance of F. Guzman, M. Xu, J. Zheng and Q. Lin.

# Exceptional electronic transport and quantum oscillations in thin bismuth crystals grown inside van der Waals materials

# Supplementary Information

Table of contents

# 1. vdW-molding technique details

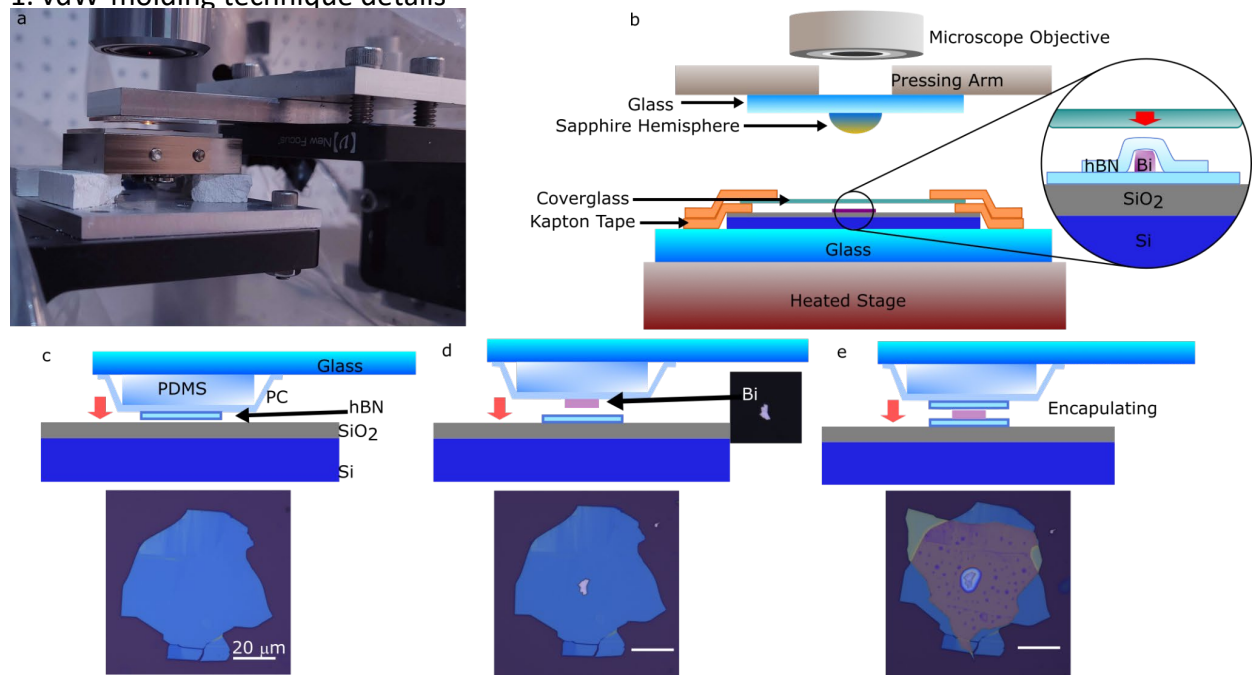

**Fig S1.** a) Photo of the microsqueezing setup. b) Diagram of the microsqueezing setup with inset of the sample stack. Setup is designed to keep the top and bottom substrates parallel during squeezing while minimizing shear forces c)-e) Diagram and optical sample image of each step to make a bismuth-hBN stack: c) transfer of bottom hBN flake onto substrate, d) transfer of starting bismuth flake on the bottom hBN, e) encapsulating bismuth with top hBN flake.

**Description of the Stacking Process:** The starting point for vdW-mold growth is to prepare bismuth micro flakes which can be easily encapsulated[1]. First, bismuth powder (Sigma-Aldrich, ~100 mesh, ≥99.99%, 140 μm in diameter) is dispersed into IPA or ethanol solutions. The mixture is drop casted on a 2 cm x 2 cm Si/SiO$_2$ chip as the bottom substrate. Another Si/SiO$_2$ chip is placed face down as the top substrate to sandwich the bismuth particles. The stacked Si/SiO$_2$ chips are placed between a hydraulic press with heaters. Typically, 1 ton of force is applied initially on 4 sets of chips. The pressure for each stack is approximately 6 MPa. Then the force is increased to 4 tons simultaneously as the stacks are heated to 204 C. The resulting hydraulic-squeezed bismuth flake is around 300 nm to 400 nm thick and is used as our starting material in the hBN-Bi-hBN stacks.

**Stacking Process:** To encapsulate bismuth between hBN, we follow the Polycarbonate (PC)-based vdW stacking technique[2]. To start, we make PC stamps from Polydimethylsiloxane (PDMS) and PC films to pick up flakes for encapsulation. PDMS (SYLGARD 184 Silicone Elastomer Kit) is mixed in a petri dish with 20 to 1 ratio of the silicone elastomer base to curing agent. PC films are made by dissolving 10% (w/v%) PC pellets (Aldrich Chemistry) in Chloroform, spreading the solution on glass slides and air drying. After the PC solution dries on the glass slide, we pick up the PC film with a hole-punched double-sided tape. On another glass slide, we cut a 5 mm x 5 mm PDMS cube and tape the PC onto the glass slide while aligning the punched hole with the PDMS cube.

We pick up the hBN flakes and hydraulic-squeezed bismuth flakes with separate PC stamps using a transfer setup. To pick up hBN flakes, we lower the PC stamp to contact with the flake at room temperature, and the stage is heated up to 110 C for 1 min. Then, we cool the stage back to room temperature while maintaining the PC contact area around the hBN flake. After the stage is cooled down, we lift the PC stamp up quickly to pick up the hBN flake. This is repeated for hydraulic-pressed bismuth flakes, but heated up to 120 C. The PC stamp will pick up multiple bismuth flakes. We choose a bismuth crystal that is less than 5 μm in size for vdW-molding.

We use a Si/SiO$_2$ chip (300 nm thick SiO$_2$ layer) 1.5 cm x 2.5 cm in size as our sample chip. The flakes are sequentially transferred onto the chip to form a hBN-bismuth-hBN stack. To transfer the flakes, we bring the flake on the PC stamp in contact with the chip, and heat the stage to above 150 C. The PC film melts at this temperature, allowing us to detach the PC film from the PDMS, leaving the PC with the flake attached to the chip. To remove the PC, the chip is submerged in Chloroform for 10 min and in IPA for 2 min. This process is repeated for the bismuth and the top hBN flake.

**Microsqueezing Setup:** The vdW-molding is performed in a home-made microsqueezing setup (Fig S1a) that can apply pressure in a ~500 μm$^2$ region locally via a sapphire hemisphere. The setup is designed to provide a simple way to bring two flat surfaces together while keeping their surfaces parallel and minimizing shear forces. The sapphire hemisphere is secured face-down on a glass slide mounted to an XYZ micromanipulator (25 mm Travel ULTRAlign Crossed-Roller Bearing Stages from Newport Co.). A microscope is integrated into the setup for in-situ observation. The sample chip is mounted onto a glass slide with kapton tape (2.5 mil thick including adhesive layer) which also acts as spacers to align the top substrate. An intermediate flexural top substrate made of a glass or sapphire coverslip (thickness 0.19 - 0.23 mm) is placed on top of the spacer kapton tape and secured with one more layer of kapton tape. The top substrate is prevented from touching the chip by the kapton tape spacers (Fig S1b) when no pressure is applied. The tape also prevents the top substrate from moving laterally during the squeezing process. The sapphire hemisphere is aligned over the stack and then pressed into the coverslip, causing it to bend down and contact the hBN-Bi-hBN stack. The purpose of the flexural coverslip is to prevent shear forces on the stack caused by uncontrolled relative lateral motion of the pressing hemisphere relative to the chip, for example, due to thermal expansion of the stage when heating.

**vdW-molding Process:** The vdW-molding process is carried out inside of a homemade glovebag with a constant flow of N$_2$ gas. After aligning the sapphire hemisphere to the hBN-bismuth-hBN stack, the sample stage is heated to 150 C (this is done solely to decrease the amount of relative thermal expansion for the final heating step). We lower the sapphire hemisphere to bend the coverslip towards the hBN-bismuth-hBN stack. When the coverslip and the stack make contact, we observe that the bismuth expands under pressure. We increase the pressure until the top and bottom substrates fully deform and come into contact around the stack and the bismuth stops expanding. After the max pressure has been applied, we heat the sample up to temperatures as high as 300 C. Typically, we observe a sudden expansion of the bismuth at 220 C, which we interpret as the melting point transition of the bismuth under the high ~1 GPa pressures. The stage is cooled down to room temperature before we lift the sapphire hemisphere and release the pressure from the stack.

**Preparing thick samples:** To prepare samples above 50 nm in thickness, bismuth is molded into hBN spacers encapsulated by hBN flakes. Spacers are lithography patterned and etched to the desirable shape. The spacer is added to the stack after the Fig S1c process. Other parameters are the same as molding bismuth between two hBN flakes.

**Exposure to Oxygen:** Under ambient conditions, bismuth has a stable native oxide layer 1-2 nm thick[3]. At high temperatures close to the melting point, bismuth oxidizes rapidly in air. To reduce the ambient oxygen concentration we use a glovebag purged with $N_2$ gas, although this only slows the rate of oxidation[4]. More significant are the protective effects of hBN encapsulation[5], which prevents catastrophic oxidation of bismuth even without the use of a glovebag. Furthermore, we find that the squeezing-induced deformation of the substrates around the bismuth-hBN stack is also effective in sealing the stack and reducing the rate of oxidation.

2. Removing the top hBN

In order to characterize the surfaces of our bismuth sample, we used a PVC pillar stamp to remove the top hBN flake[6]. First, we made a small block of PDMS with 10 μm x 10 μm x 20 μm pillar structures. This is done by pouring the PDMS solution into a SU-8 mold with an array of holes. After the PDMS is cured, we cut out the pillar PDMS piece and place it on a glass slide. We secure the PVC film (Riken Wrap from Riken Fabro Co., Ltd) over the pillar onto the glass slide using hole-punched double-sided tapes, similar to making PC stamps. The glass slide is mounted face-down on the transfer arm. First, we align the pillar on top of the top hBN flake. Then, we approach the top hBN flake slowly with the PVC-PDMS pillar as we heat the stage up to 55 C, around which PVC becomes sticky. Depending on the humidity, the temperature might need to be increased to 70 C or higher[7]. Once the pillar touches the top hBN flake, only a small contact area is made between the top hBN flake and the PVC. Keeping the stage at ~55 C, we slowly lift up the PVC-PDMS pillar stamp with piezo-motors, which peels the top hBN flake away from the bismuth.

## 3. Overview of hBN-molded bismuth results

In this section we provide additional data on the hBN-molded bismuth samples we have produced. We measured 40 vdW-mold samples in AFM that produced ultraflat bismuth crystals with thicknesses ranging from 5 nm to 150 nm.

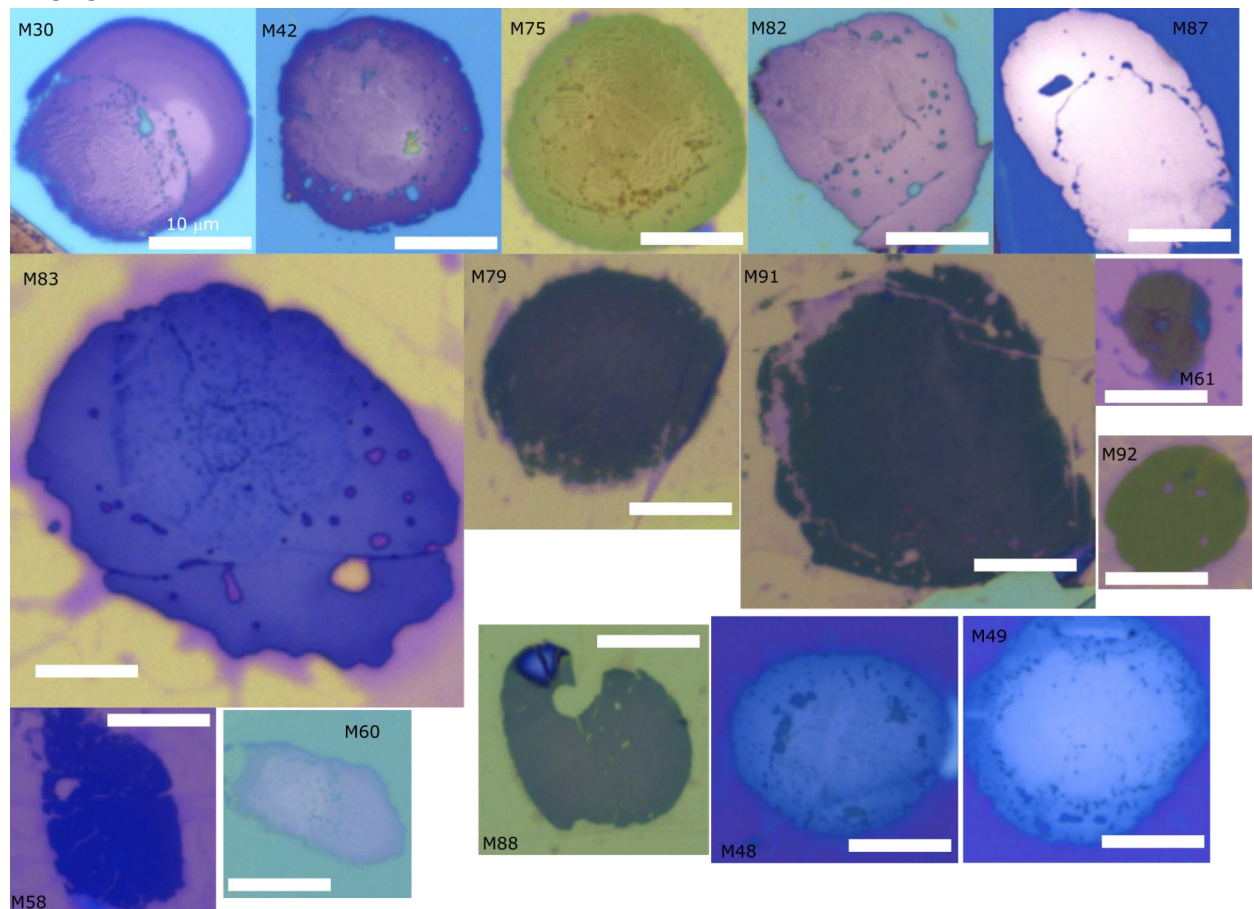

**Fig S2. Optical images of vdW-molded bismuth samples.** For samples M60 and M87 the top hBN flakes have been removed. Scale bar is the same for all the images 10 µm.

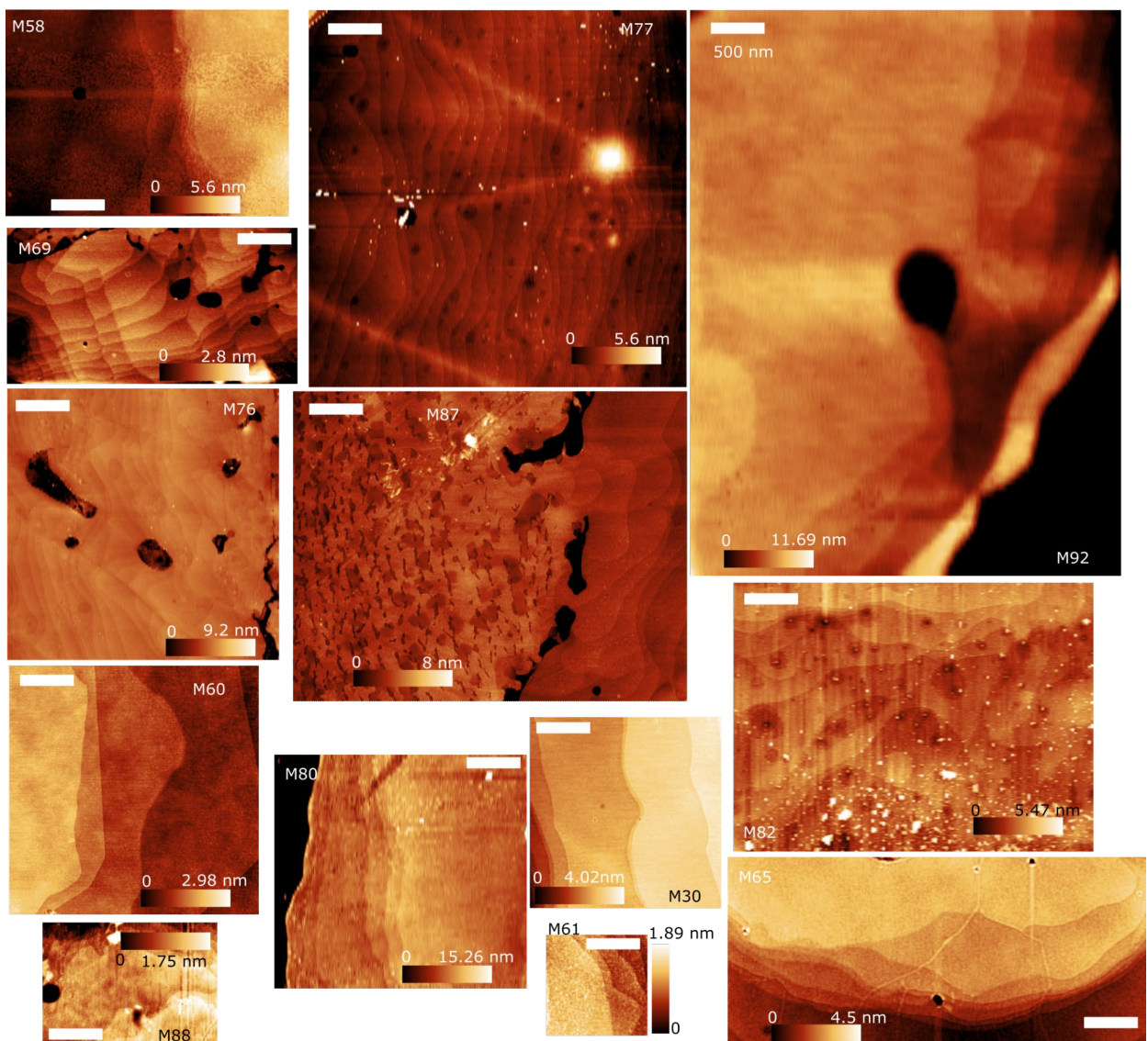

**Fig S3. AFM scans of the hBN-molded bismuth surfaces showing various flat terrace structures.** All samples have the top hBN removed, except for M92. Clear layered terraces are visible in M92 through the thin hBN layer.

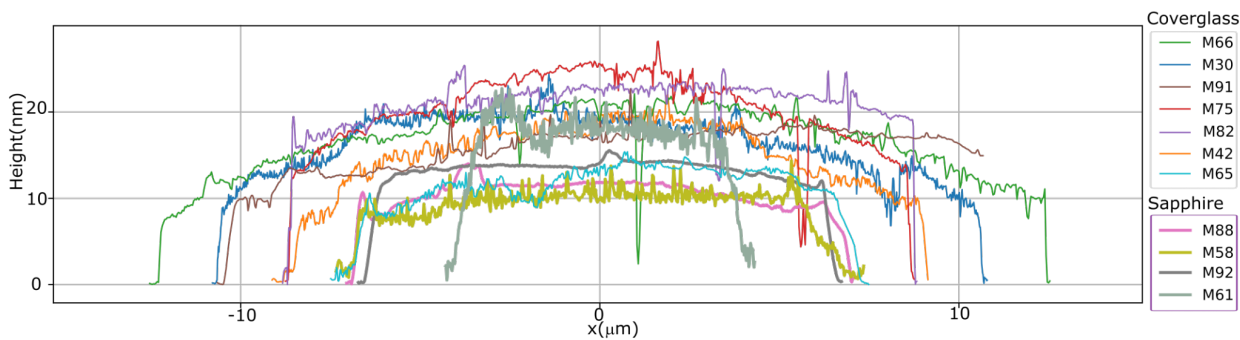

**Fig S4. AFM line profiles of hBN-molded bismuth samples squeezed by glass or sapphire top substrates showing a dome shape.** Squeezing with sapphire as the top substrate leads to overall thinner bismuth with a flatter profile.

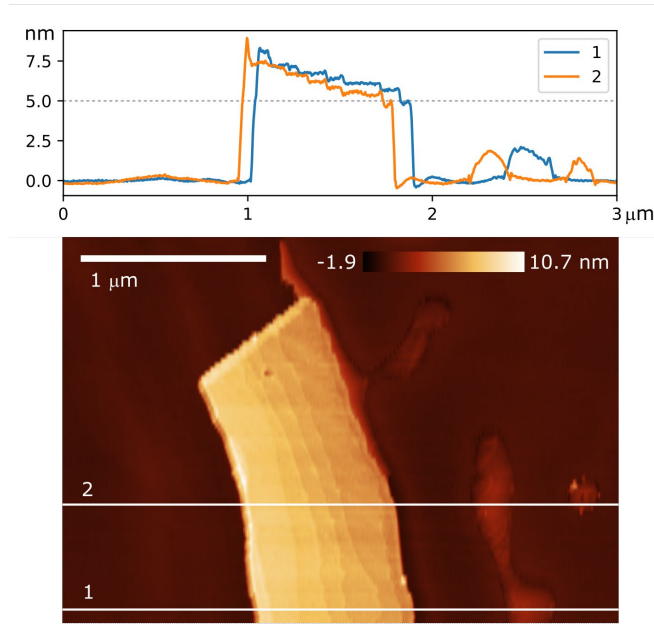

**Fig S5.** AFM topography and line profile of a vdW-molded bismuth showing a 5 nm thick terrace.

**Table 1: Table of vdW-molded bismuth samples.**
(Only thin samples molded without hBN spacers are shown here.)

| Sample Name | Max thickness (nm) | Min thickness (nm) | Diameter (µm) | Substrate | Max Terrace width (µm) |
|---|---|---|---|---|---|
| M60 | 15 | 12 | 18 | Glass | 2.1 |
| M69 | 16 | 12 | 11 | Glass | 0.4 |
| M65 | 17 | 12 | 13 | Glass | 1.0 |
| M48 | 17 | 12 | 20 | Glass | |
| M42 | 19 | 9 | 21 | Glass | |
| M77 | 21 | 12 | 26 | Glass | 0.9 |
| M30 | 21 | 6 | 21 | Glass | 2.2 |
| M91 | 22 | 13 | 32 | Glass | |
| M66 | 22 | 9 | 27 | Glass | 0.7 |
| M79 | 25 | 13 | 17 | Glass | 1.0 |
| M75 | 26 | 16 | 24 | Glass | 0.3 |
| M87 | 26 | 20 | 27 | Glass | 1.2 |
| M76 | 27 | 13 | 23 | Glass | 0.4 |
| M59 | 33 | 18 | 9 | Glass | 0.6 |
| M49 | 39 | 13 | 24 | Glass | 1.0 |
| M58 | 10 | 7 | 17 | Sapphire | 1.8 |
| M88 | 11 | 9 | 18 | Sapphire | 2.4 |
| M80 | 11 | 11 | 26 | Sapphire | 1.3 |
| M92 | 15 | 11 | 14 | Sapphire | 3.8 |
| M61 | 20 | 12 | 9 | Sapphire | 1.2 |

## 4. Squeezing with other substrates

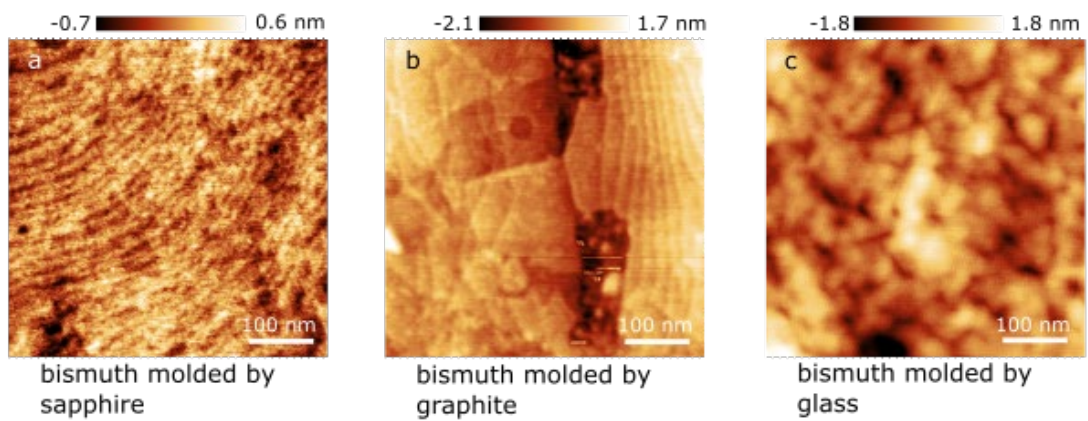

**Fig S6. AFM topography of bismuth molded under pressure between different substrates.**

**Table 2: RMS roughness of different molding substrates and the corresponding molded bismuth**

| Substrate | RMS roughness of the substrate | RMS roughness of the molded bismuth |
|---|---|---|
| Sapphire | 0.63 Å | 2.1 Å |
| Graphite[8] | 0.23 Å | 1.0 Å |
| Glass | 1.3 Å | 5.3 Å |

To study the imprinting effect of substrates besides hBN on the molded bismuth, we did AFM analysis (Fig S6) on bismuth molded directly between sapphire wafers (University Wafer 2" sapphire substrate wafer, 100 +/- 25 µm <0001> +/- 1°), graphite and microscope cover glass.

The average root mean square (RMS) roughnesses of the molding substrates and the molded bismuth are listed in Table 2. The roughness of the bismuth follows the trend of the substrates. Graphite-molded bismuth shows terrace structure similar to the hBN-molded bismuth.

## 5. TEM and EBSD characterization

**Sample Preparation and TEM information:** We transfer the vdW-molded ultrathin bismuth onto a TEM grid (200 nm Silicon Nitride Membrane TEM grid with 2.5 µm holes from Ted Pella, inc.) via the PVC transfer method explained in the previous section[6]. Samples are measured in a JEOL-2100F TEM with Schottky-type field emission electron source with 80-200 kV acceleration voltage. Point to point resolution is 0.27 nm and lattice resolution is 0.1 nm.

**Additional TEM images and diffraction measurements**

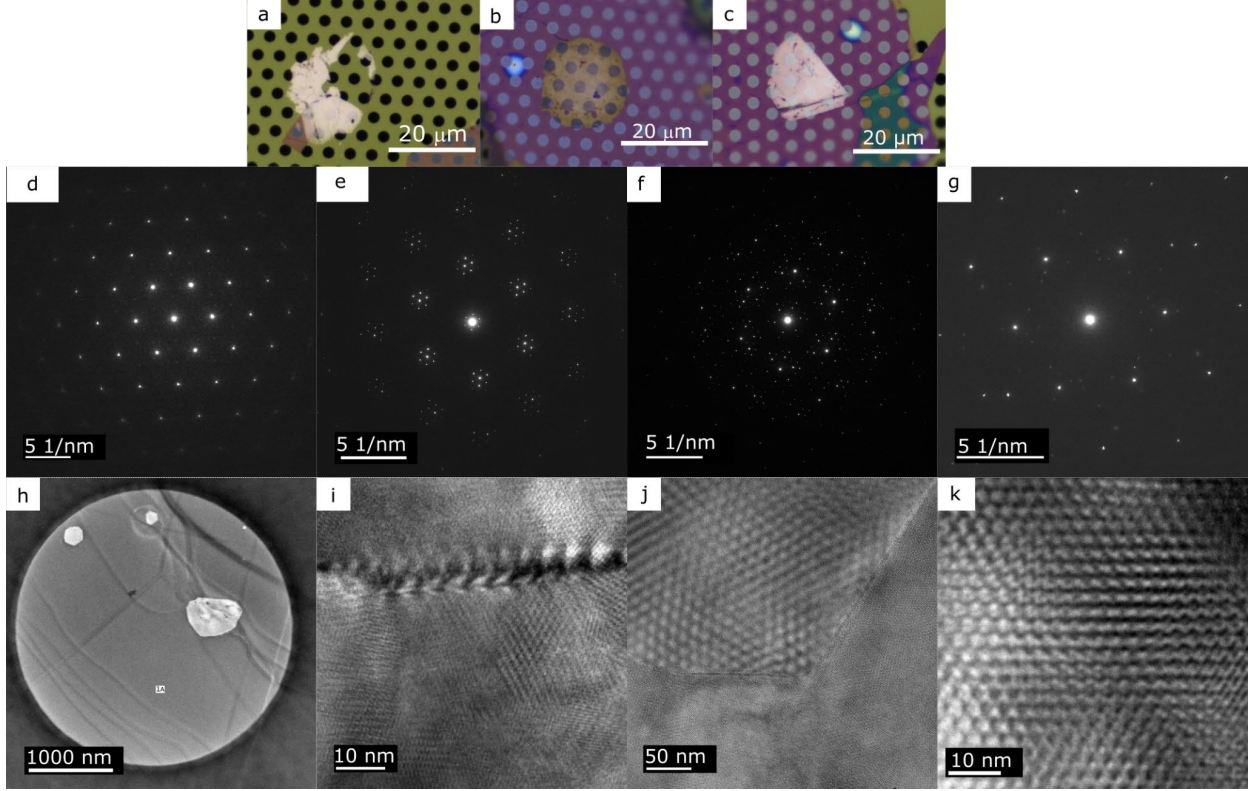

**Fig S7. 3 samples measured in TEM all exhibited rhombohedral lattice structure with the 111 axis oriented out of plane.** a)-c) Optical image of vdW-molded bismuth samples on TEM grids. a) vdW-molded bismuth with hBN flakes removed, b) vdW-molded bismuth encapsulated between hBN flakes and c) vdW-molded bismuth with bottom hBN flake. d) Selected area diffraction (SAED) of bismuth (70 nm) molded between the hBN stack. e) SAED of hBN-molded bismuth (30 nm) with the bottom hBN. f) SAED of hBN encapsulated bismuth (11 nm thick bismuth). g) SAED of 23 nm thick hBN-molded bismuth with both top and bottom hBN flakes removed. h) TEM image of hBN-molded bismuth showing terraces similar to terraces observed in AFM. i) TEM image of grain boundaries on hBN-molded bismuth without hBN. j) TEM image of moire patterns formed from different bismuth domains between hBN flakes. k) TEM image of moire pattern formed by bismuth on bottom hBN.

**Calculation of lattice constant from diffraction images:** In the selected area diffraction pattern for bismuth, we observe the ($1\underline{1}0$) family of diffraction spots (Fig 2f). We describe the bismuth lattice using the rhombohedral unit cell with parameters from the hexagonal close-packed structure: $a$ is the in-plane lattice constant and $c$ is the out-of-plane lattice constant[9]. The reciprocal space lattice vectors of the bismuth rhombohedral unit cell can be described by the following, where $\hat{x}$, $\hat{y}$ and $\hat{z}$ are cartesian coordinate vectors.

$$b_1 = \frac{-2\pi}{a}\hat{x} - \frac{2\pi\sqrt{3}}{3a}\hat{y} + \frac{2\pi}{c}\hat{z}$$

$$b_2 = \frac{2\pi}{a}\hat{x} - \frac{2\pi\sqrt{3}}{3a}\hat{y} + \frac{2\pi}{c}\hat{z}$$

$$b_3 = \frac{2\pi\sqrt{3}}{3a}\hat{y} + \frac{2\pi}{c}\hat{z}$$

From our data, we measure the magnitude of the reciprocal space lattice vector from ($1\bar{1}0$) to ($\bar{1}10$) diffraction peaks. The measured magnitude is $0.434$ Å$^{-1}$, resulting in a value of $a = 4.62$ Å. This value is the same as is observed in real space images and is close to the bulk bismuth lattice constant of $4.54$ Å.

**Electron beam backscattering (EBSD)**

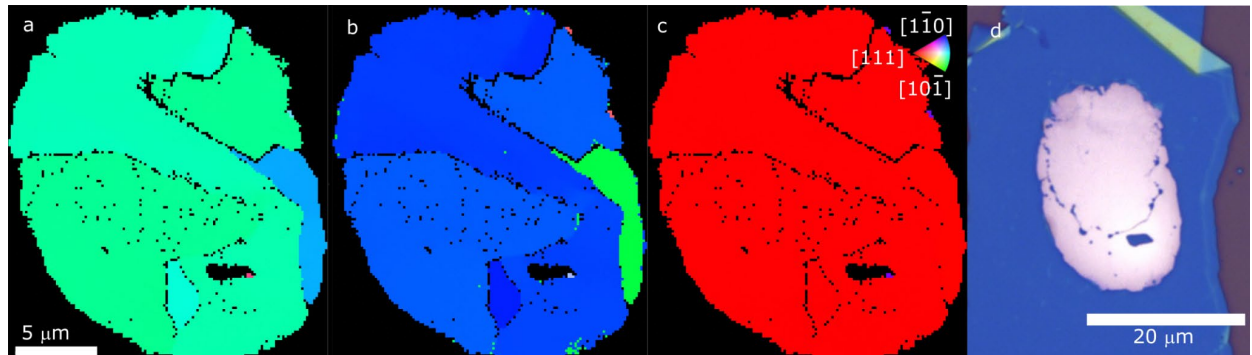

**Fig S8.** a)-c) Electron beam backscattering (EBSD) inverse pole figures (IPF) of hBN-molded bismuth sample. a) Y-axis IPF, b) X-axis IPF, c) Z-axis IPF, d) Optical image of hBN-molded bismuth used for EBSD (top hBN removed).

The sample is measured in a Tescan GAIA3 Scanning Electron Microscope (SEM) operating at 20 kV. EBSD data (Fig S8a-c) is processed through an open source tool (MTEX) in Matlab. The data shows that all the domains of this sample are in the (111) orientation. We performed EBSD on 4 samples in total, 3 molded by hBN and 1 molded by SiO2. All samples exhibited the bismuth rhombohedral crystal structure with a 111 orientation.

## 6. DFT calculations

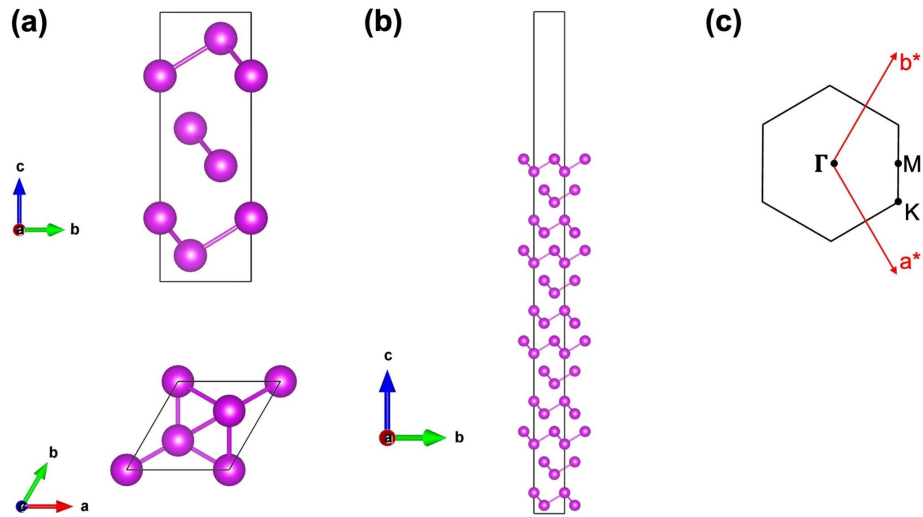

**Fig S9.** (a) Lattice structure of the bulk bismuth (111) for the top and side views. (b) Lattice structure of 12 bilayer bismuth with two surfaces. (c) The 2D first Brillouin zone of bismuth (111).

Our first-principles calculations are performed with the projector-augmented wave pseudopotentials[10,11] and the generalized gradient approximation of Perdew-Burke-Ernzerhof[12] using Vienna Ab initio Simulation Package[13] code. An energy cutoff of 450 eV and a 10 × 10 × 1 Monkhorst–Pack k-point grid is used[14]. The structure is optimized until the atomic forces are smaller than 0.01 eV/Å. The vacuum layer is larger than 19 Å to ensure decoupling between neighboring nanostructures. For the vdW corrections, we used the DFT-D3 method[15,16].

Fig S9a shows the lattice structure of the bulk bismuth, with the (111) orientation aligned with the c-axis, corresponding to the bulk band structures (gray) in Fig. 3c from the main text. The lattice constants are a=4.57 Å, c=11.75 Å. The Bi bond length is 3.10 Å (purple bonds in Fig S9a). The intra-bilayer height is 1.63 Å and inter-bilayer spacing is 2.29 Å. Then, we build a 12 bilayer (111)-oriented thin film with two (111) surfaces, as shown in Fig. S1b. After relaxation, the average of the intra bilayer height is 1.50 Å and inter bilayer spacing is 2.13 Å. The 2D reciprocal lattice is shown in Fig. S9c.

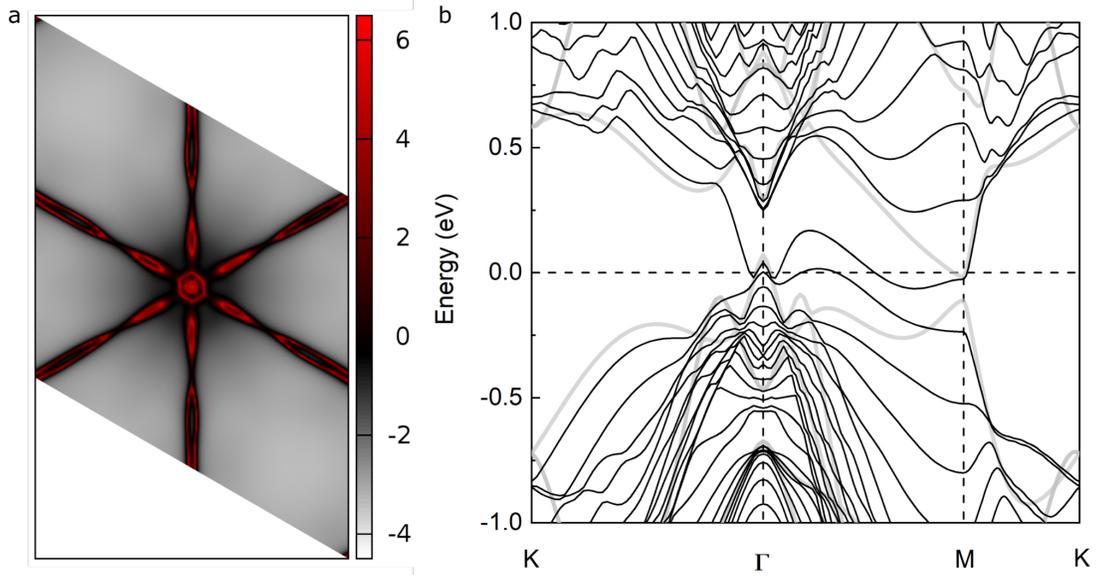

**Fig S10.** (a) Fermi surface at kz=0, calculated using WannierTools[17] with a dense 201 × 201 × 1 k mesh, based on the tight-binding Hamiltonian obtained by employing maximally localized Wannier functions (MLWFs) method using WANNIER90[18] with initial projections to Bi-p orbitals. (b) Band structure with Fermi surfaces at 0 eV. This plot is used as the basis of the Fermi surface schematic in the main text Figure 4.

## 7. Raman characterization

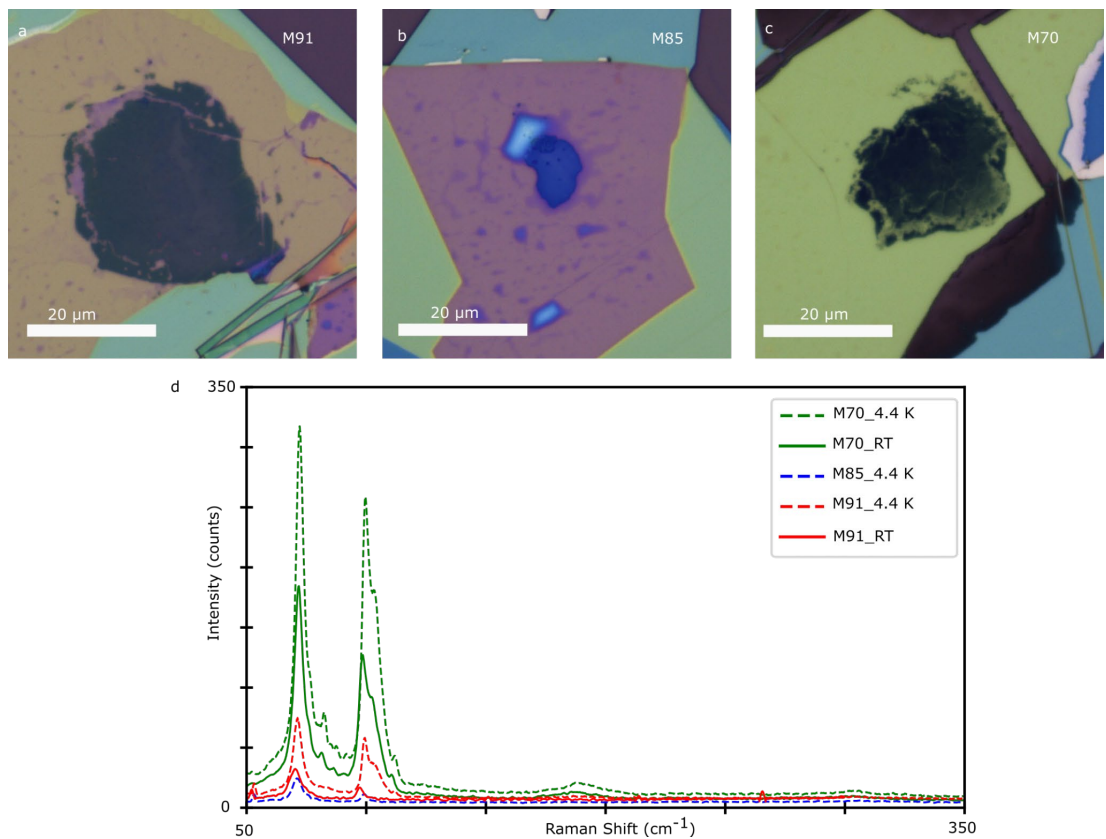

**Fig S11.** a)-c) Optical image of bismuth hBN-encapsulated samples. d) Raman data of M70, M91 at room temperature (solid lines) and M70, M85, M91 at 4.4 K (dash lines).

Raman data are taken in a Horiba LabRAM HR Evolution with an ultra-low frequency module (532 nm laser). The hBN-molded bismuth samples are on Si chips with 300 nm $SiO_2$. M91 and M85 are encapsulated stacks of hBN-molded bismuth. At room temperature, we observed two peaks in M70 at 71.9 $cm^{-1}$ and 98.6 $cm^{-1}$ (M91 at 70.3 $cm^{-1}$ and 97.3 $cm^{-1}$). These values closely match the frequencies of the normal modes observed in bulk rhombohedral bismuth at 71 $cm^{-1}$ ($E_g$) and 98 $cm^{-1}$ ($A_{1g}$)[19]. In M91 and M85, we do not observe any bismuth oxide related Raman modes. In M70, we observe a faint peak at 185.9 $cm^{-1}$ which matches the Bi-O stretching mode at 185 $cm^{-1}$ [4]. During the molding process of M70, most of the melted bismuth flowed out of the bottom hBN flake, leaving a higher concentration of the bismuth oxide from the source flake (Fig S9c). We conclude that our samples are rhombohedral bismuth encapsulated between hBN flakes with minimal oxidation.

## 8. Model of squeezing limits and molecular dynamics simulations

**Simple Continuum Model**

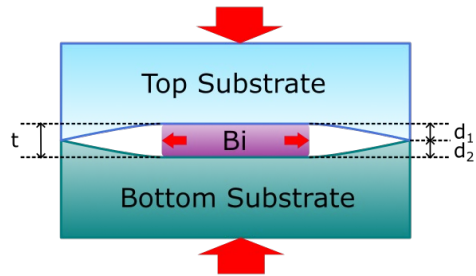

**Fig S12. Depiction of substrate deformation during the squeezing process.**

The squeezing limit model considers a balance of the pressure to squeeze the liquid bismuth due to its surface tension $P = 2S/t$ with the pressure to indent the substrate $P = 2rEd/\pi r^2 = 2Ed/\pi r$, where r is the droplet radius, E is the substrate's elastic modulus, d is the depth of indentation, and S is the the difference in surface energy per unit area between the bismuth-hBN and hBN-hBN interfaces (i.e. the relative surface tension). Assuming identical top and bottom substrates, the limiting condition occurs when $t = 2d$, i.e. the bismuth is fully indented symmetrically into both substrates. From this condition we derive the thickness limit $t = \sqrt{2\pi S_T\, r/E}$. In the case where the substrates are different (ex: Silicon on the bottom, glass on top), then we can relate the depths of indentation into each substrate with the total thickness $t = d_1 + d_2$, where $d_1$ and $d_2$ are the depth of indent for the bottom and top substrates, respectively. The resulting derived equations are $d_1^2 = \pi r S_T / E_1\, (1 + E_1/E_2)$ and $d_2^2 = \pi r S_T / E_2\, (1 + E_2/E_1)$.

Table 3 shows calculated values for the limiting thickness under various conditions. For these estimates, we use the bare bismuth surface tension value of S = 0.388J/m² [20]. Note that this neglects the interaction between the bismuth and hBN, and also neglects that growing the surface area of the bismuth requires breaking hBN-hBN interlayer bonds at the periphery of the laterally expanding droplet. The hBN-hBN cohesion energy is measured to be 0.326 J/m² [21], potentially nearly doubling the true value for S.

The primary results in this paper are for Si:Glass and Si:Sapphire squeezing, with predicted limits of 15 nm and 10.6 nm, respectively, using the above equations. This matches well the range of thicknesses we observe in our actual squeezings. Switching to dual Sapphire substrates would reduce the predicted thickness to 9 nm, and reducing the droplet radius by 4x would give an additional 2x reduction in thickness. Thinner squeezing may be achievable by circumventing the substrate deformation trapping, for example, by concentrating pressure into smaller regions.

**Table 3: Limiting thickness value at max pressure for r = 10 μm droplet following the simple deformation model**

| Squeezing substrates | thickness (nm) |
|---|---|
| Si:Si (E=170GPa) | 12 |
| **Si:Glass** (170GPa:70GPa) | 15 |
| **Si:Sapphire** (170GPa:300GPa) | 10.6 |
| Sapphire:Sapphire (300GPa) | 9 |

**Molecular Dynamics Simulations of Squeezing**

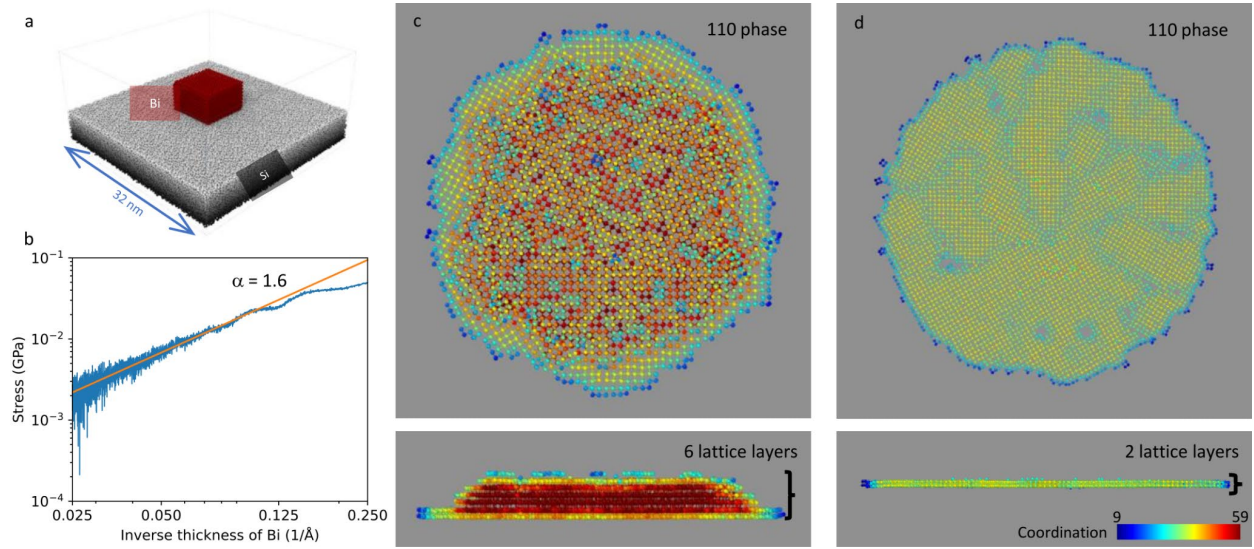

**Fig S13. Molecular dynamics simulation of bismuth squeezed by silicon.** a) 3D view of the simulation setup, consisting of a bismuth block on a silicon slab. b) Blue: Simulation data showing uniaxial compressive stress as a function of the inverse thickness of the squeezed bismuth in the liquid state (temperature 550 K). Orange: Power-law fitting to data with exponent alpha. c) Resulting bismuth crystal after melt-growth while squeezing with a 2 nm gap with a Si (111) substrate. Top view and side view (upper and lower panels, respectively). d) Same simulation as in (c) but with a 0.75 nm gap between squeezing substrates. The colors of the atoms represent the sum of coordination numbers from 1st to 5th nearest neighbors (c-d).

To further study the limits of the squeeze method of producing ultrathin crystals, we performed molecular dynamics (MD) simulations of a nanodroplet of bismuth squeezed between planer indenter and silicon substrates. In the simulations, we find that it is possible to achieve crystals thinner than 1 nm when using starting bismuth material that is 301 nm$^3$ in volume and at applied uniaxial compression of 21 MPa (Fig S13). At this ultrathin limit, the bismuth is sensitive to the substrate interaction, in this case showing a (110)-orientated crystal when squeezed by (111)-orientated silicon (Fig S13c&d) and an amorphous phase when squeezed by (100)-orientated silicon (not shown). Note that silicon was chosen as the squeezing substrate to simplify the simulation as compared to the more complex dynamics of a hBN crystal.

**Details of Molecular Dynamics Simulations**

Molecular dynamics simulations were performed using the open-source code LAMMPS[22]. The simulation consisted of a bismuth block compressed by a silicon substrate. The dimensions of the bismuth are 77.16 Å, 78.62 Å and 49.68 Å along x, y and z cartesian coordinates, respectively. The (111) plane of bismuth is perpendicular to the vertical direction, i.e., along the z-axis. The dimensions of the silicon were 325 Å, 325 Å and 55 Å, respectively. Periodic boundary conditions were applied along all three dimensions. Modified embedded-atom method (MEAM) was used to compute non-bonded interactions of bismuth[23]. Interactions between silicon atoms were modeled with a 3-body Stillinger-Weber (SW) potential[24]. The interactions between bismuth atoms and silicon atoms on the interface were described by the Lennard-Jones (LJ) potential, in which the well depth ε and size-parameter σ are 0.019996 eV and 3.225 Å, respectively[25]. The temperature was maintained at 300 K by a Langevin thermostat with the damping parameter set to 1 ps. A planar indenter with a force constant equal to 10 eV/Å$^3$ was used to compress the model from above the bismuth block with a constant velocity 0.1 Å/ps. The bottom of the silicon substrate was fixed to prevent movement. After the compression, the model was heated to the melting point (550 K) of bismuth for 50 ps. The planer indenter was maintained above the bismuth droplet to avoid the escape of bismuth atoms. The bismuth droplet was then cooled down to 300K at a constant cooling rate 1 K/ps. The atomic structures were visualized by the program OVITO[26].

## 9. Device fabrication details

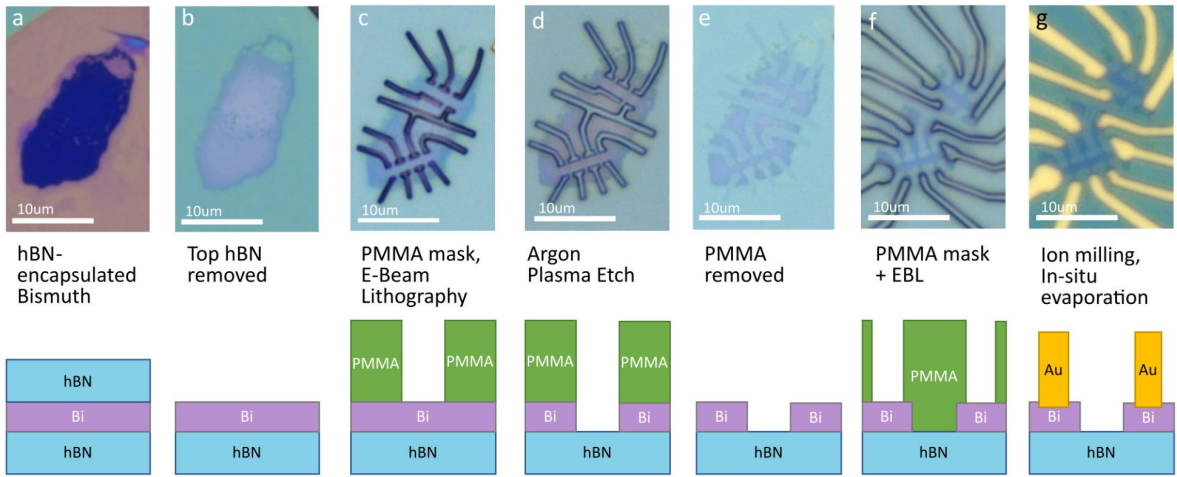

**Fig S14. Optical image and cross-sectional schematics of the fabrication of the bismuth transport devices.**

The 8 nm and 13 nm devices are fabricated from vdW-molded bismuth samples with the top hBN removed, and the main transport channel between the voltage tabs is from one terrace of uniform thickness. The fabrication procedure for the open-face devices are shown in Fig S14.

We begin with a vdW-molded bismuth encapsulated between hBN on a Si/SiO$_2$ chip.
First, the top hBN flake is removed via the PVC method[6] explained in a previous section. Then the bismuth surface topography is measured through AFM, allowing us to locate the regions containing large flat terraces with uniform thicknesses. In order to create Hall-bar shaped devices, we spin coat 1 μm of PMMA (Microchem, 950 A7) onto the chip by spinning at 2000 rpm for 2 min, and performed e-beam lithography (FEI Magellan 400 XHR SEM) to make the PMMA into an etch mask. The bismuth is then etched into the designed shape via Ar plasma etch (Plasma-Therm Reactive Ion Etcher Model 790) with the following parameters: 300 W power, 401 V DC bias voltage, Ar flow rate 30 sccm, process pressure 115 mTorr for 45 s. It is followed by $O_2$ plasma etch of 70 W power, 227 V DC bias voltage, O2 flow rate 9.9 sccm, process pressure 70 mTorr for 10 s. This $O_2$ plasma etch is done in order to break the cross-linked PMMA. After removing the PMMA in acetone, the chip is spin coated with 1 μm of PMMA and made into a mask for electrodes via e-beam lithography. Since the bismuth is exposed to air, a thin oxide layer forms at its surface and results in a large contact resistance. We remove the oxide layer by Ar ion milling (Intlevac Nanoquest 1) with 400 V beam voltage, 30 mA beam current, 80 deg beam angle, 80 V acceleration voltage for 5 s at 10 C, and immediately deposit 5 nm Cr and 15 nm Au on top without exposing the bismuth to atmosphere (Telemark 249, built-in within the Ion Mill). Finally, after taking out the bismuth sample, we evaporate 5 nm Cr and 100 nm Au (Temescal CV-8) on the bismuth and lift off the PMMA in acetone.

A similar fabrication process is also used to produce devices from encapsulated vdW-molded bismuth samples without removing the top hBN. Note, when the top hBN is thin, we are still able to observe the bismuth terraces through the hBN by AFM (as shown in Fig S3). In the bismuth geometry defining etch (Fig S14d), we performan extra step of $SF_6$ plasma etch to etch away the top hBN before etching the bismuth with Ar plasma. The etching parameters are 30 W power, 15 V DC bias voltage, $SF_6$ flow rate 10 sccm, process pressure 100 mTorr for 15 s. In the step where we define the electrodes (Fig S14g), we also did $SF_6$ plasma etch to remove the top hBN in the unmasked region before ion milling the bismuth oxide.

**Role of bismuth oxide:** The vdW-molded bismuth is protected from air during the growth process. However, the device nanofabrication process exposes the bismuth to some oxygen. Bismuth forms a self-limiting oxide layer about 1-3 nm in thickness as we discussed in Section 1, and we can see signs of a crystalline oxide layer in our TEM measurements. As such, likely all of our electronic measurements of clean quantum oscillations in the bismuth surface states are coming from electrons near this oxide layer. Given the high quality of the resulting transport data and the agreement with STM and ARPES studies, we conclude that this oxide layer does not significantly degrade transport. Instead, it seems that the surface roughness is the most important factor for thin bismuth transport. These observations are also supported by DFT calculations, which show that the bismuth surface state is barely affected by the presence of an oxide layer[27].

## 10. Additional transport data

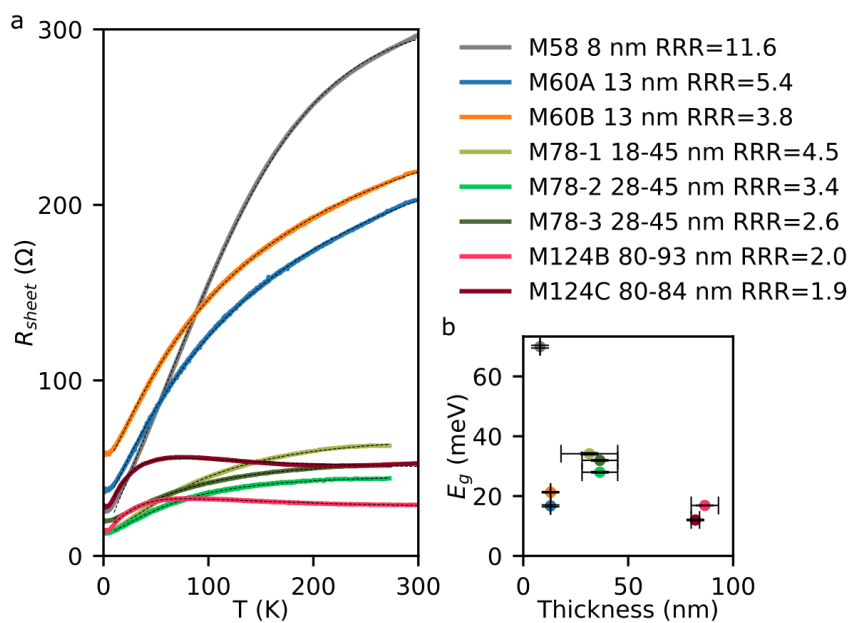

**Fig S15. Temperature-dependent transport measurements of various devices, plotted in designated colors. a,** Sheet resistance as a function of temperature. **b,** Fitted bulk gap as a function of device thickness.

Fig S15 shows the temperature dependence resistance measurements on various devices. We fit these data to a resistance model which consists of metallic conduction and activated conduction in parallel, where $R_0$ is the residual resistance, a is the metallic dependence prefactor, $G_1$ is the semiconducting dependence prefactor k, $E_g$ is the bulk gap, $k_B$ is the boltzmann constant[28].

$$R = [(R_0 + aT)^{-1} + G_1 exp(-\frac{E_g}{2k_B T})]^{-1}$$

The fitted curves are plotted as dashed lines on top of the sheet resistance in Fig S15a. The fitted parameter $E_g$ is plotted as a function of device thickness in Fig S15b.

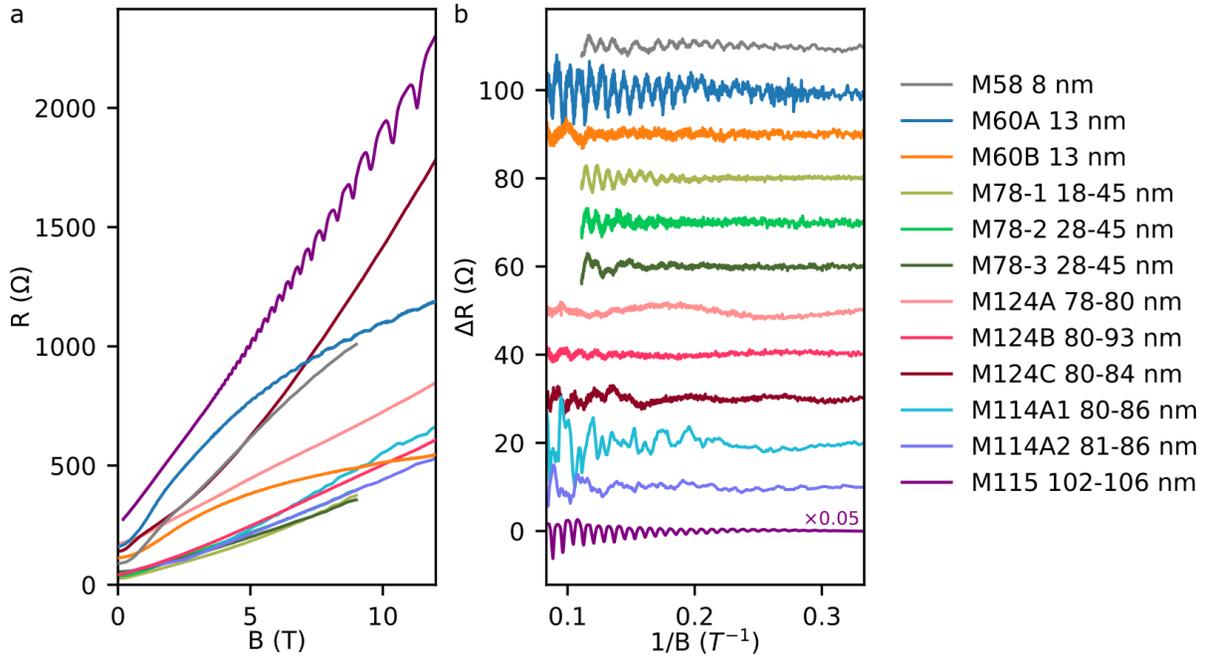

**Fig S16. Field-dependent transport measurements from various devices. a,** Resistance as a function of magnetic field. **b,** Quantum oscillations in $\Delta R(1/B)$ under 12T, calculated by subtracting a smoothed background.

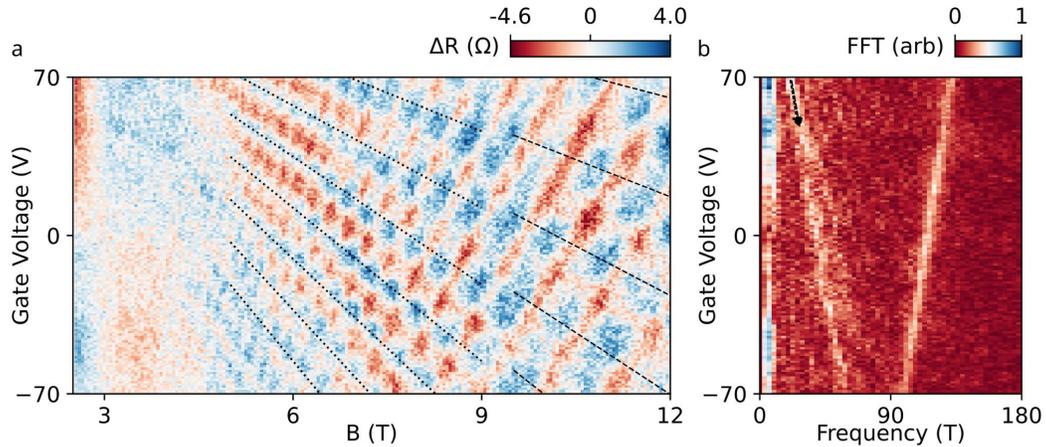

**Fig S17. Left**, Quantum oscillations measured as a function of gate voltage and magnetic field for the rough 13 nm sample (1.1 μm x 2.1 μm) at 1.5K. Same device as appears in main text Figure 3. The amplitude of the quantum oscillation is much smaller in comparison to the flatter sample. **Right**, FFT calculation of the quantum oscillation of the sample. Unlike the flat sample, we observe only one dominant hole oscillation.

## 11. Two-carrier model fit

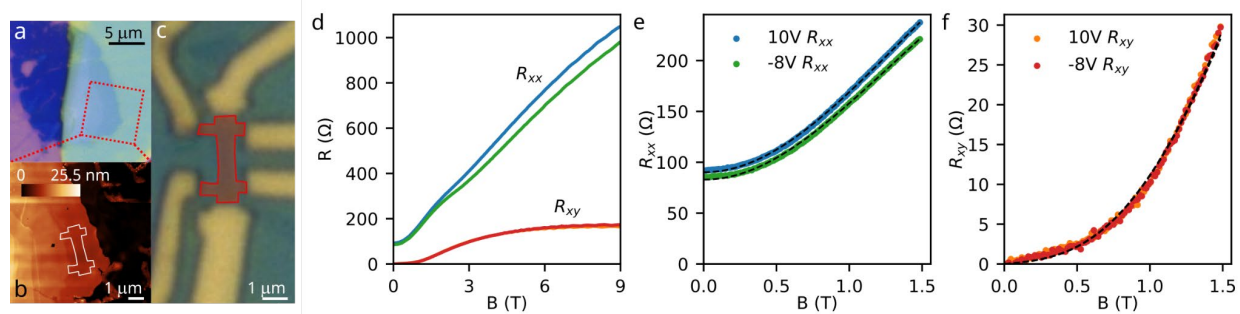

**Fig S18. Magnetic field-dependent measurement of the longitudinal resistance (Rxx) and the transverse resistance (Rxy) of a 8nm flat device at two different back gate voltages.** a, Optical image of the bismuth where this device is fabricated. b, AFM image of the area marked out in a. c, Optical image of the fabricated device. d, $R_{xx}$ and $R_{xy}$ plotted as a function of magnetic field up to 12 T. e, Two-carrier model fit result under 1.5 T plotted on top of $R_{xx}$. f, Two-carrier model fit result under 1.5 T plotted on top of Rxy.

We are able to apply the semimetal two-carrier model[29] on the longitudinal ($R_{xx}$) and transverse resistance ($R_{xy}$) (Fig S18d) of M58, a flat 8 nm device (Fig S18a-c). We corrected the $R_{xy}$ data by subtracting the 0-field resistance, because we did not obtain resistance in the negative field to symmetrize the data.

$$\rho_{xx} = \frac{1}{e} \frac{n\mu_n + p\mu_p + (n\mu_p + p\mu_n)\mu_n\mu_p B^2}{(n\mu_n + p\mu_p)^2 + (p-n)^2\mu_n^2\mu_p^2 B^2}, \quad \rho_{xy} = \frac{1}{e} \frac{(p\mu_p^2 - n\mu_n^2)B + (p-n)\mu_n^2\mu_p^2 B^3}{(n\mu_n + p\mu_p)^2 + (p-n)^2\mu_n^2\mu_p^2 B^2}$$

Where n and p are the carrier density of electrons and holes, $\mu_n$ and $\mu_p$ are the corresponding mobility. We fit the $R_{xx}$ and $R_{xy}$ data under 1.5 T to the model, and the fit result is plotted on top of the data in Fig S18e and f, and the fitted parameters are listed below.

**Table 4: Carrier densities and mobilities from the two-carrier model fit**

| Gate Voltage (V) | n (cm$^{-2}$) | p (cm$^{-2}$) | $\mu_n$ (cm$^2$/Vs) | $\mu_p$ (cm$^2$/Vs) |
|---|---|---|---|---|
| 10 | 7.63e+12 | 1.72e+13 | 1.22e+04 | 8.67e+03 |
| -8 | 7.99e+12 | 1.84e+13 | 1.26e+04 | 8.84e+03 |

—